\documentclass[10pt]{article}

\usepackage{times}
\usepackage{graphicx}
\usepackage{multicol}

\makeatletter
\def\@normalsize{\@setsize\normalsize{10pt}\xpt\@xpt
\abovedisplayskip 10pt plus2pt minus5pt\belowdisplayskip
\abovedisplayskip \abovedisplayshortskip \z@
plus3pt\belowdisplayshortskip 6pt plus3pt
minus3pt\let\@listi\@listI}

\def\subsize{\@setsize\subsize{10pt}\xipt\@xipt}

\def\section{\@startsection {section}{1}{\z@}{1.0ex plus 1ex minus .2ex}{.2ex plus .2ex}{\large\bf}}

\def\subsection{\@startsection {subsection}{2}{\z@}{.2ex
plus 1ex}{.2ex plus .2ex}{\subsize\bf}}

\def\keywords{\vspace{.6em}
    \if@twocolumn
    \small{\itshape Keywords\/}\bfseries---$\!$%
    \else
    \small{\itshape Keywords\/}\bfseries---$\!$%\end{center}\quotation\small
    \fi}
\def\endkeywords{\vspace{0.6em}\par\if@twocolumn\else\endquotation\fi
    \normalsize\rm}

\makeatother
\begin{document}

\date{}

\newcommand{\mytitle}{\vskip -5mm \hskip -7mm By chance is not enough: preserving relative density through non uniform sampling}
%Multiple Views for Web search Result Visualization
\title{\Large\bf \mytitle}

\author{Enrico Bertini, Giuseppe Santucci\\
        {\small Dipartimento di Informatica e Sistemistica - Universit\`{a} di Roma ``La Sapienza''}\\
        {\small Via Salaria, 113 - 00198 Roma, Italy - \{bertini, santucci\}@dis.uniroma1.it}
}

% I don't know why I have to reset thispagestyle, but
% otherwise get page numbers
%\pagestyle{empty}
%\thispagestyle{empty}

%\begin{multicols}{2}
\twocolumn[\maketitle]

\subsection*{\centering Abstract}
{\em

Dealing with visualizations containing large data set is a
challenging issue and, in the field of Information Visualization,
almost every visual technique reveals its drawback when
visualizing large number of items. To deal with this problem we
introduce a formal environment, modeling in a virtual space the
image features we are interested in (e.g, absolute and relative
density, clusters, etc.) and we define some  metrics able to
characterize the image decay. Such metrics drive our automatic
techniques (i.e., not uniform sampling) rescuing the image
features and making them visible to the user. In this paper we
focus on 2D scatter-plots, devising a novel non uniform data
sampling strategy able to preserve in an effective way relative
densities.}

\begin{keywords}
visual clutter, metrics, non-uniform sampling.
\end{keywords}

%\pagestyle{empty}
%%%%%%%%%%%%%%%%%%%%%%% Introduction %%%%%%%%%%%%%%%%%%
\section{Introduction}
\label{sec:intro}

Visualizing large data sets results, very often, in a cluttered
image in which a lot of graphical elements overlap and many pixels
become over plotted, hiding from the user the main image visual
features.

We deal with this problem providing a formal framework to measure
the amount of decay resulting from a given visualization, then we
build, upon these measures, an automatic non uniform sampling
strategy that aims at reducing such a degradation. We focus on a
very common visual technique, 2D scatter-plots,
analyzing the loss of information derived by overlapping pixels.

In this paper we improve and extend some preliminary results
presented in \cite{BerSan04}, defining a formal model that
estimates the amount of overlapping elements in a given area and
the remaining free space. These pieces of information give an
objective indication of what is eventually visualized on the
physical device; exploiting such measures we can estimate the
quality of the displayed graphic devising techniques able to
recover the decayed visualization.

To eliminate the sense of clutter, we employ a low grain non
uniform sampling technique dealing with the challenging issue of
devising the \emph{right} amount of sampling in order to preserve
the visual characteristics of the underlying data. It is quite
evident, in fact, that a too strong sampling is useless and
destroys the less dense areas, while a too light sampling does not
reduce the image clutter.  The formal model we discuss in the
paper gives precise indications on the right amount of data
sampling needed to produce a representation preserving the most
important image characteristics, i.e., relative densities that are
one of the main clues the user can grasp from 2D scatter-plots.

The contribution of this paper is twofold: (1) it presents a
formal model that allows for defining and measuring data density
both in terms of a virtual space and of a physical space (e.g., a
display) and (2)  it defines a novel automatic non uniform
sampling technique driven by some  metrics defined above
the previous figures.

The paper is structured as follows: Section \ref{sec:relwork}
analyzes related works, Section \ref{sec:modeling} describes the
model we use to characterize clutter and density,
formalizing the problem and introducing the metrics we are
interested in, Section \ref{sec:sampling} describes our non
uniform sampling technique, Section \ref{sec:discussion} discusses
the results obtained applying our techniques to a real data set,
and, finally, Section \ref{sec:conclusions} presents some
conclusions, open problems and future work.

%%%%%%%%%%%%%%%%%%%%%%%%%%%%%%% Related Work %%%%%%%%%%%%%%%%%%%%%%%%%%%%%%%%%%%%%

\section{Related Work}
\label{sec:relwork}

This paper deals with issues concerning metrics for Information Visualization and
techniques to address the problem of overlapping pixels and visual
clutter in computer displays. In the following we
illustrate the research proposals closer to our approach and their relationship with our work.
\vspace{0.8em}

\subsection{Metrics for Information Visualization}
As expressed in \cite{need-metrics}, Information Visualization
needs metrics able to give precise indications on how effectively
a visualization presents data and to measure its \emph{goodness}.
Some preliminary ideas have been proposed considering both formal
measurements and guidelines to follow.

Tufte proposes in \cite{tufte} some measures to estimate the
quality of 2D representations of static data. Measures like the
\textit{lie factor}, that is the ratio of the size of an effect as
shown graphically to its size in the data, are examples of first
attempts to systematically provide indications about the quality
of the image displayed. Tufte's proposal however applies to paper
based 2D visualizations and does not directly apply to interactive
computer-based images. Brath in \cite{metrics-brath}, starting
from Tufte's proposal, defines new metrics for static digital 3D
images. He proposes metrics such as \textit{data density} (number
of data points/number of pixels) that resemble Tufte's approach
together with new ones, aiming at measuring the visual image
complexity. The \textit{occlusion percentage}, for example, has
connections with our work. It provides a measure of occluded
elements in the visual space suggesting to reduce such a value as
much as possible. These metrics are interesting and are more
appropriate for describing digital representations. However, as
stated by the author, they are still immature and need
refinements.

While the above metrics aim at measuring a general
goodness or at comparing different visual systems, our aim is to
measure the accuracy of the visualization, that is, how well it
represents the characteristics hidden inside data. They present
some similarities with past metrics but operate at a lower level
dealing with pixels and data points, providing measures that can
directly be exploited to drive corrective actions. It is worth to
note that, on the contrary of the above proposals, we will show
how the suggested metrics can be exploited \textit{in practice} to
take quantitative decisions about corrective actions and enhance
the current visualization.\vspace{0.8em}

\subsection{Dealing with overlapping pixels and clutter}
The problem of eliminating visual clutter and overlapping pixels
to produce intelligible graphics has been addressed by many
proposals.

\textit{Jittering} as stated in \cite{point-occlusion}, is a widely adopted
technique that permits to make apparent pixels that naturally map
into the same position into the screen. The idea is to slightly
change the position of overlapping points in order to render them
all visible. Similarly, space-filling pixel-based techniques
\cite{keim-gridfit} distribute data points along predefined curves
to avoid overlapping pixels, shifting them to
positions that are as close as possible to the original one.

\textit{Transparency} is also an interesting technique to overcome
occlusion and reduce clutter, both in in 3D \cite{transparency}
and 2D \cite{million-vis} visualizations.  However, when dealing
with pixel-based visualizations it is not possible to convey
transparency at the level of single pixels, though it is useless.

\textit{Constant density visualization} \cite{vida}\cite{vida-2}
is an interesting technique to deal with clutter. Exploiting the
idea of generalized fisheye views \cite{generalized-fisheye}, it
consists in giving more details to less dense areas and less
details to denser areas, allowing the screen space to be optimally
utilized and to reduce clutter. The problems with this approach
are that it requires the user to interact with the system, the
overall trend of data is generally lost, and some distortions are
introduced.

\textit{Sampling} is used in \cite{dix-ellis} to reduce the
density of visual representation. As the authors state, if the
sampling is made in random way, the distribution is preserved and
though it is still possible to grasp some useful information about
data correlation and distributions, permitting \textit{``to see
the overall trends in the visualization but at a reduced
density''}. Even if interesting, this idea is not free of
drawbacks. In particular, when the data present particular
distributions, i.e., the data set has both very high and very low
density areas, choosing the right amount of sampling is a
challenging task. Depending on the amount of sampling two problems
can arise: 1) If the sampling is too \textit{strong} the areas in
which the density is under a certain level become completely
empty; 2)If the sampling is too \textit{weak} the areas with
higher densities will still look all the same (i.e., completely
saturated) and consequently the density differences among them
will be not perceived by the user. A first proposal in this direction is in \cite{BerSan04}, where an automatic uniform sampling technique is presented, able to compute the optimal sampling ratio w.r.t. some quality metrics.

Our approach differs from  the above proposals for three main
aspects:
\begin{itemize}
\item it provides a sound model for defining in both a virtual and physical space several metrics intended specifically for digital images;
\item it provides, on the basis of the above figures, some \emph{quantitative} information about the image decay;
\item it exploits such numerical results for automatically computing \emph{where}, \emph{how}, and \emph{how much} to sample preserving, as much as possible, a certain visual characteristic.
\end{itemize}

%%%%%%%%%%%%%%%%%%%%%%%% Modeling Visual Density and Clutter  %%%%%%%%%%%%%%%%%%
%intro of section
\section{ Modeling Visual Density and Clutter}
\label{sec:modeling}

\begin{figure}[t]
\centering
\includegraphics[width=0.35\textwidth]{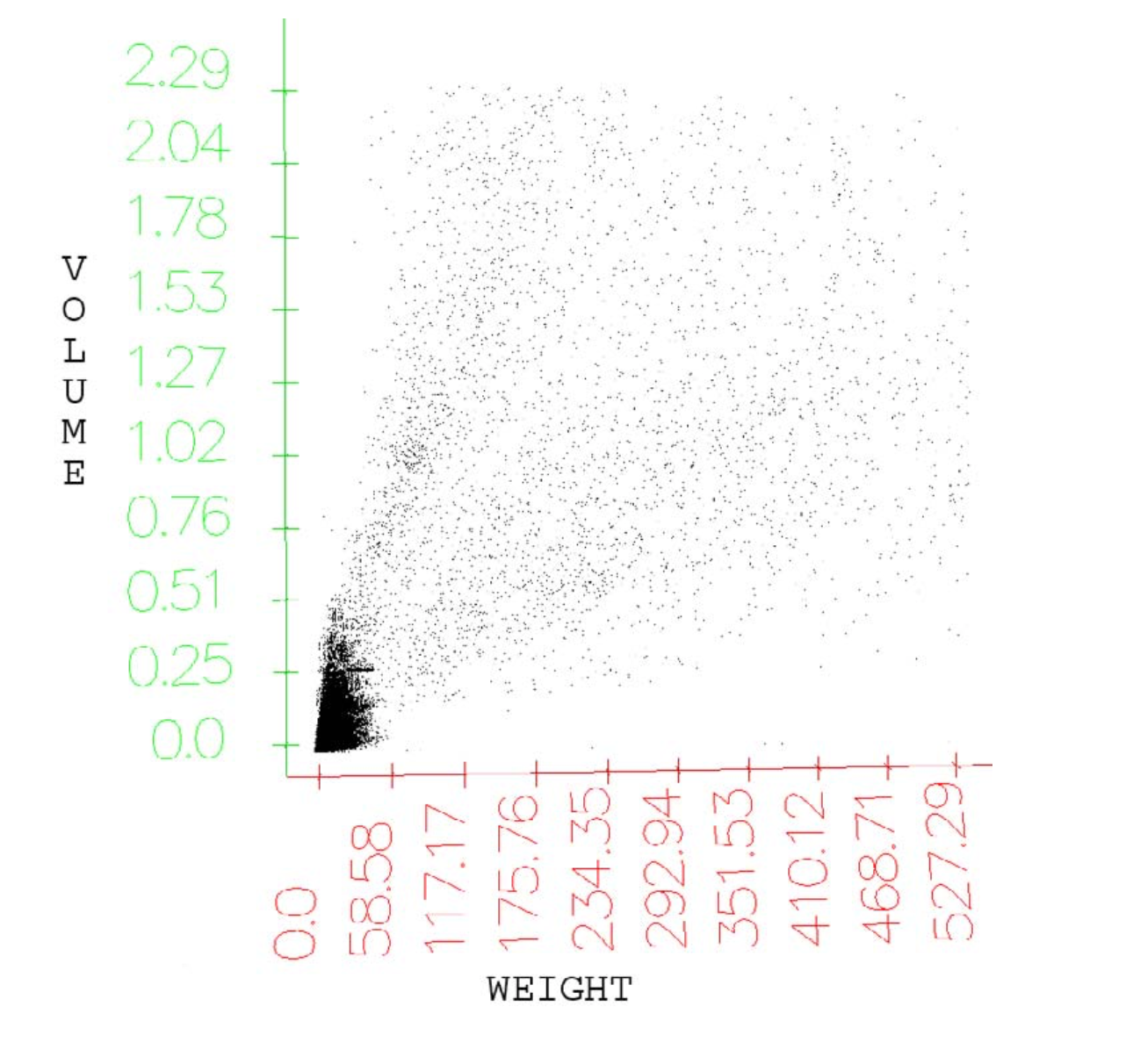} \caption{{\em
Plotting mail parcels}} \vspace{-0mm} \label{fig:parcel}
\end{figure}

In this section we present the formal framework that aims at
modeling the clutter produced by over-plotting data. Some preliminary issues about the matter are in \cite{BerSan04}; here we show a refinement of that results.

We consider a 2D space in which we plot elements by
associating a pixel to each data element mapping two data attributes on the  spatial coordinates.
As an example, Figure~\ref{fig:parcel} shows about 160,000 mail parcels
plotted on the X-Y plane according to their weight (X axis) and
volume (Y axis). It is worth noting that, even if the number of plotted items is little, the area close to the origin is very
crowded (usually parcels are very light and little), so a great
number of collisions is present in that area: the most crowded area contains more that 50,000 (about 30 \%) of the whole dataset compressed in less than 1 \% of the whole screen.

%what the functions do
Exploiting well known results
coming from the calculus of probability, we derive a function that
estimates the amount of colliding points and, as a consequence,
the amount of free available space. More formally, two points are
in collision when their projection is on the same physical pixel. In order
to derive such a function, we imagine to \textit{toss} $n$ data
points in a  random way on a fixed area of $p$ pixels. This assumption is quite
reasonable if we conduct our analysis on {\em small} areas.

%parameters
To construct such functions we use a probabilistic model based on
the parameters just described, that here we summarize for the sake
of clarity:

\begin{itemize}
\item $n$ is the number of points we want to plot;
\item $p$ is the number of available pixels;
\item $k$ is the number of collisions;
\item $d$ is the number of free pixels.
\end{itemize}

The probability of having {\em exactly} $k$ collisions  plotting
$n$ points on an area of $p$ pixels, $Pr(k, n, p)$, is given by
the following function:
\begin{displaymath}
\begin{array}{cl}
\frac{PERM[{p \choose n-k}{n-k+k-1 \choose k}]}{p^{n}} &
 \mbox{if $n\leq p$ and $ k \in [0,n-1] $} \\
 \ & \mbox{or $n>p$ and $k \in [n-p,n-1]$}\\
 \ & \ \\

0 & \mbox{if $n>p$ and $ k \in [0,n-p] $}\\
\end{array}
\end{displaymath}

The function is defined only for $k<n$, because it
 is impossible to have more collisions than plotted points. Moreover, it is easy to understand that in some cases the probability is
equal to zero: if $ n>p $, because of we are plotting more points
than available pixels, we
 must necessarily have some collisions. For example, if we have an area of $8 \times 8$ pixels and we plot 66 points, we must necessarily have at least 2 collisions, so $Pr(0, 66, 64)=0$ and $Pr(1, 66, 64)=0$.

%spiegazione della formula
The basic idea of the formula is to calculate, given $p$ pixels
and $n$ plotted points, the ratio between the number of possible
cases showing  {\em exactly} $k$ collisions and the total number of
possible configurations.

The latter is computed  considering all
the possible ways in which it is possible to choose $n$ points among
 $p$ pixels, i.e., selecting
$n$ elements from a set of $p$ elements allowing repetitions
(dispositions with repetitions: $p^{n}$).

Calculating the \textit{\# config with exactly k collisions} is
performed in three steps. First we calculate all the possible ways
of selecting $n-k$ non colliding points from $p$ pixels
(combinations \textit{without} repetitions: ${p \choose n-k}$).
After that, for each of such combinations, we calculate all the
possible ways of hitting $k$ times one or more of the $n-k$ non
colliding points in order to obtain exactly $k$ collisions, that
corresponds to selecting $k$ elements form a set of $n-k$ elements
\textit{with} repetitions (combinations \textit{with} repetitions:
${n-k+k-1 \choose k}$). Finally, because of we are interested in
all the possible dispositions, we need to count the permutations
(PERM) of these combinations. Unfortunately, because of the
variable number of duplicates (e.g., it is possible to have k
collisions hitting k+1 times the same pixel $p_{i}$, or k times
$p_{i}$ and two times pixel $p_{j}$, or k-1 times $p_{i}$, two
times pixel $p_{j}$, and two times pixel $p_{k}$ and so on) we
were no able to express such permutations by a close formula.

%obtaining the graphs
>From the above expression we derived, through a C program, a
series of functions (see Figure~\ref{fig:graph}) showing the
behavior of the observed area as the number of plotted points
increases. More precisely, we compute the available free space $d$
(Y axis, as percentage w.r.t. $p$), the mean of colliding elements
$k$ (Y axis, as percentage w.r.t. $n$) for any given number of
plotted points $n$ (X axis,  as percentage w.r.t. $p$). For
example, if we have an area of 64 pixels, the graph tell us that
plotting 200\% (128) of $p$ points  will produce an average of
56.7\% (72.5) collisions. On the other hand, if we plot 128 points
having 72.5 collisions we can compute the free pixels $d$, as
$d=64-(128-72.5)=8.5$ (13.3\%).

The behavior of the functions is quite intuitive: as the number of
plotted points $n$ increases the percentage of collisions
increases as well while the free space decreases; roughly
speaking, we can say that over plotting four times the screen
results in a totally saturated display (1.6\% of free space).

Such functions can tell us how much we are saturating the space or, as a
more complex possibility, the way in which the display is able to
represent relative densities and how much to sample the data to
guarantee a prefixed visualization quality. This result is
exploited in the next section and we clarify it through an
example. Assume that we are plotting $n$ points on the area
$A_{1}$ turning on $p_{1}$ pixels and $2n$ points on the area
$A_{2}$ turning on $p_{2}$ pixels. In principle, the user should
perceive area $A_{2}$ as containing more (i.e., twice as many)
points as area $A_{1}$. Because of collisions, $p_{2}\leq
2p_{1}$ and as $n$ increases  the user initially looses the
information that area $A_{2}$ contains twice as many points as
$A_{1}$  and for greater values of $n$ the user is not able to
grasp any difference between $A_{1}$ and $A_{2}$. As a numerical
example, if we plot 64 and 128 points on two $8\times 8$ areas , the pixels
turned on in the two areas will be $40.55$ and
$55.43$, so the ratio of displayed pixels is only 1.36.
In order to preserve the visual impression that area $A_{2}$ contains
twice as many points as $A_{1}$ accepting a decay of 20 per cent
we have to sample the data (64 and 128 points) as much as 50 per
cent resulting in 32 and 64 points that, once plotted, turn on
25.32 and 40.55 pixels, i.e., a ratio of 1.6 (20 per cent of
decay).

\begin{figure}
\centering
\includegraphics[width=0.45\textwidth]{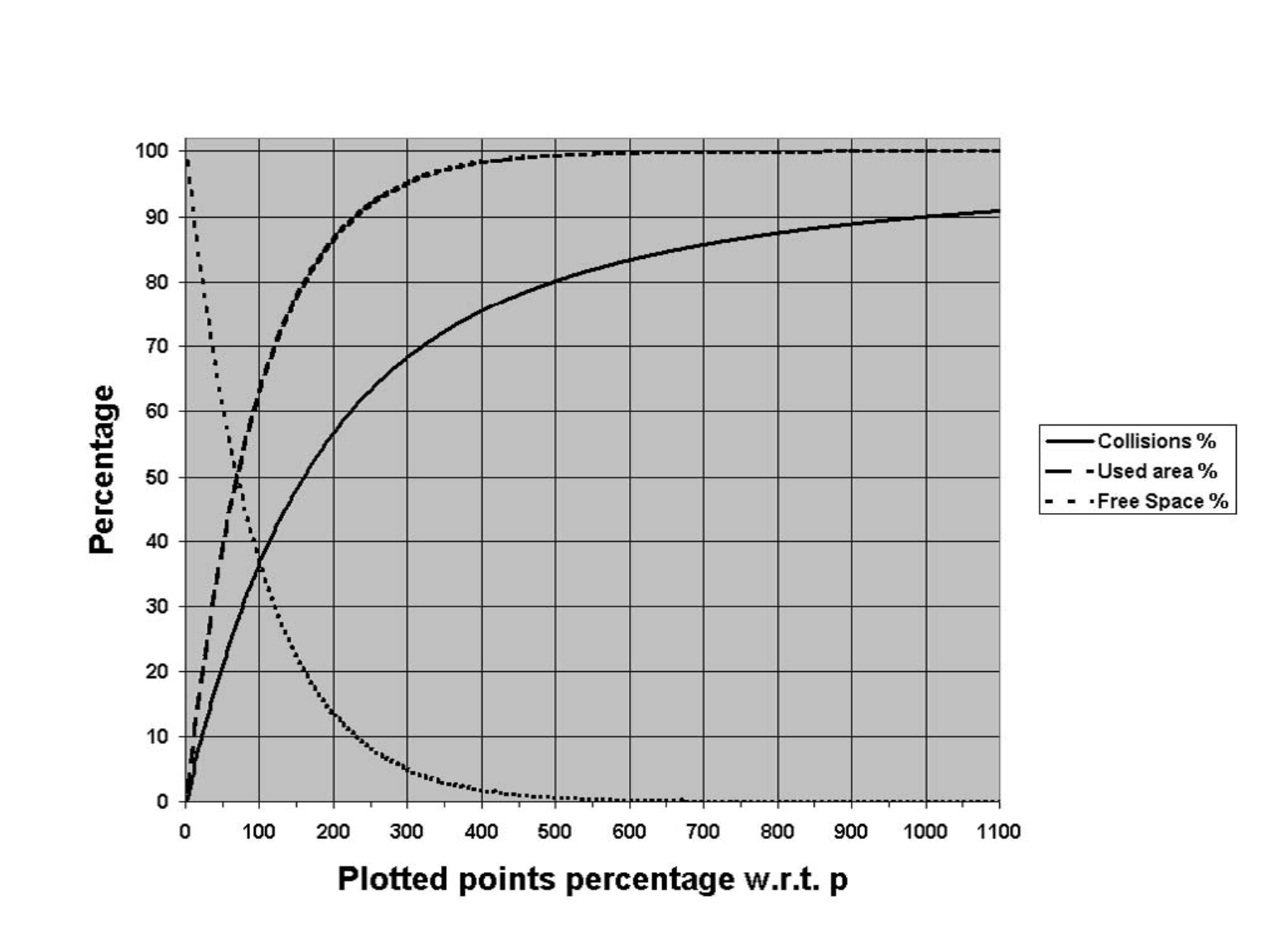}
\caption{{\em Colliding elements percentage}}
\vspace{-5mm} \label{fig:graph}
\end{figure}

\vspace{0.8em}
\subsection{Data densities and represented density}
\label{sec:metrics} The previous results give us a way to control
and measure the number of colliding elements. Before introducing
our optimization strategy, we need to clarify our scenario and to
introduce new figures and definitions.

We assume the image is displayed on a rectangular area
(measured in inches) and that small squares of area A divide the
space in $m\times n$ \textit{sample areas} ($SA$) where density is
measured. Given a particular monitor, resolution and size
affect the values used in calculations. In the
following we assume that we are using a monitor of 1280x1024
pixels and size of 13''x10.5''. Using these figures we have
1,310,720 pixels and if we choose $SA$ of side $l=0,08$
inch, the area is covered by 20.480 (128x160) sample areas whose
dimension in pixel is $8 \times 8$. We consider small areas
because of it makes the uniform distribution assumption quite realistic.

For each $SA_{i,j}$, where $1 \leq i \leq m$ and $1 \leq j \leq
n$, we calculate two different densities : \textit{real data
density} (or, shorter, data density) and \textit{represented
density}.

\emph{Data density} is defined as $d_{i,j}=\frac{n_{i,j}}{A}$ where
$n_{i,j}$ is the number of data points that fall into sample area $A_{i,j}$.
For a given visualization, the set of data densities
is finite and discrete. In fact, if we plot a number $n$ of data
elements into the display, each $SA_{i,j}$ assumes a value $d_{i,j}$
that is within the finite and discrete set of values: $0,
\frac{1}{A}, \frac{2}{A}, \dots , \frac{n}{A}$. In general, for
any given visualization, a subset of these values will be really
assumed by the sample areas. For each value we can compute the
number of sample areas in which that value is present and an
histogram showing the distribution of the various data densities
can be computed. For example, if we plot 100 data points into an
area of 10 sample areas, we could have the following
configuration: 3 sample areas with 20 data points, 2 sample areas
with 15 data points, 2 sample areas with 5 data points.

\emph{Represented density} is defined as $rd_{i,j}=\frac{p_{i,j}}{A}$
where $p_{i,j}$ is the number of distinct active pixels that fall
into $SA_{i,j}$. The number of different values that a sample area
can assume is heavily dependent on the size of sample areas. If we
adopt sample areas of size 8x8 pixels, as described before, the
number of different not null represented densities is $64$. Thus, we can
represent at most $64$ different represented density values.
It is quite obvious that, because of collisions, $rd_{i,j}\leq d_{i,j}$.

Using the above definitions we devised an effective set of quality
metrics whose complete discussion, however, is out of the scope of
this paper (see \cite {BerSan04} for a practical use of these
quality metrics for uniform sampling strategies).

The above metrics, together with the statistical results give us
the means to devise an automatic non uniform sampling technique
described in the next section.

\section{Non uniform sampling}
\label{sec:sampling} In \cite{BerSan04} a uniform sampling
strategy has been presented, showing its ability in improving an
image readability. Applying the same amount of sampling to the
whole image is quite straightforward but presents several
drawbacks. As an example, it is quite obvious that sampling areas
presenting very low data density is useless and potentially
dangerous. Moreover it is quite evident that the most important
clues a user can grasp from 2D scatter-plots are
\textbf{differences} in densities and our opinion is that a non
uniform sampling can preserve in a more efficient way such
differences.

\begin{figure*}[t]
\centering
\includegraphics[width=0.95\textwidth]{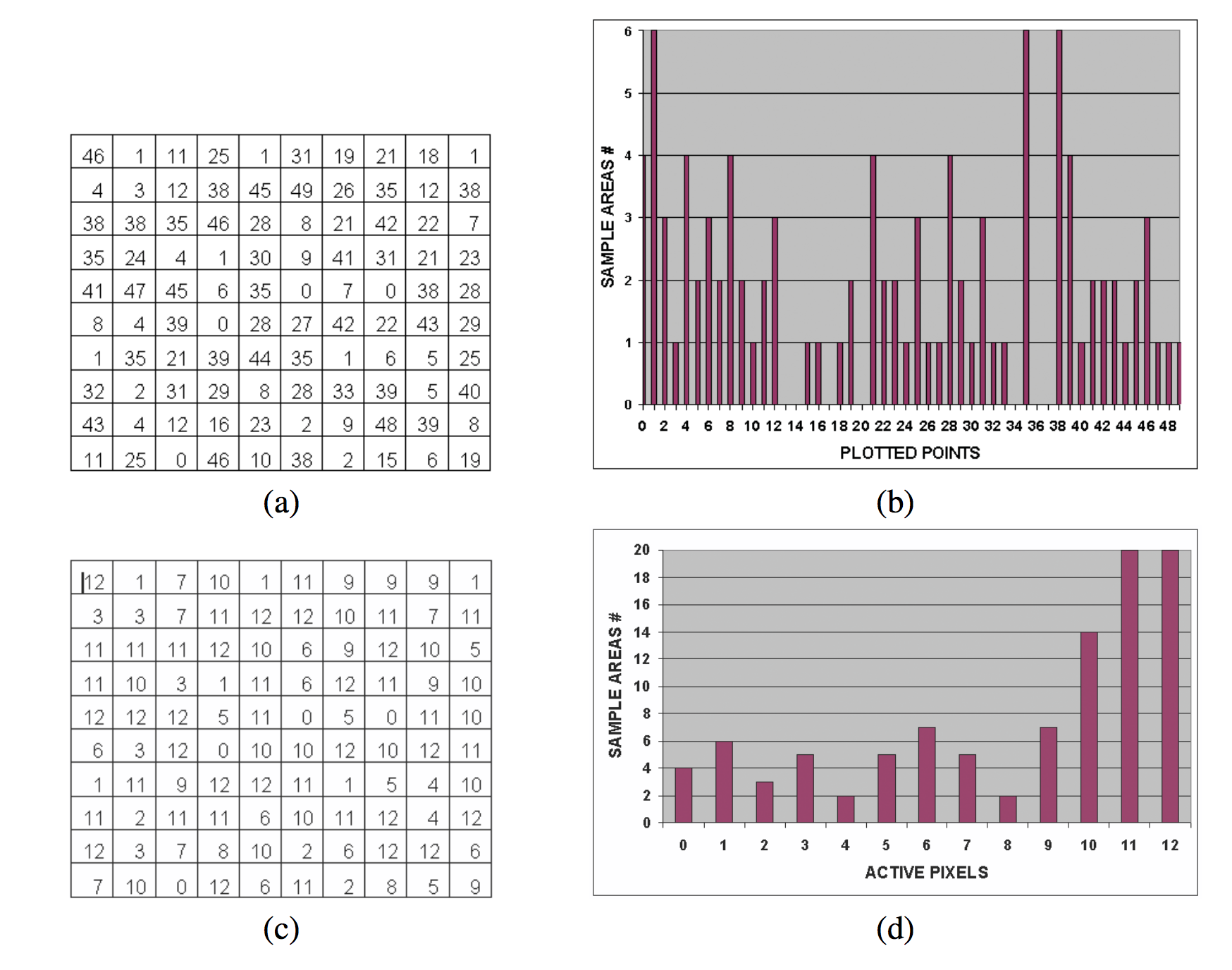}
\caption{\em A screen area made of 100 sample areas: real (a) and represented (c) data densities}
\label{no-sampling}
\vspace{-5mm} \label{fig:graph}
\end{figure*}

%\begin{figure*}[t]
%\vskip -20mm
%\begin{center}
%\begin{tabular}{c@{\hspace{1cm}}c}
%  \resizebox{4.5cm}{!}{\includegraphics{data-densities.eps}} &
%  \resizebox{6.5cm}{!}{\includegraphics{data-densities-hist.eps}} \\
%  (a) & (b) \\
%  \resizebox{4.5cm}{!}{\includegraphics{repres-densities-c0.eps}} &
%  \resizebox{6.5cm}{!}{\includegraphics{repres-densities-c0-hist.eps}} \\
%  (c) & (d) \\
%\end{tabular}
%\caption{A screen area made of 100 sample areas: real (a) and represented (c) data densities}
%\label{no-sampling}
%\vspace{-0mm}
%\end{center}
%\end{figure*}

\begin{figure*}[t]
\centering
\includegraphics[width=0.95\textwidth]{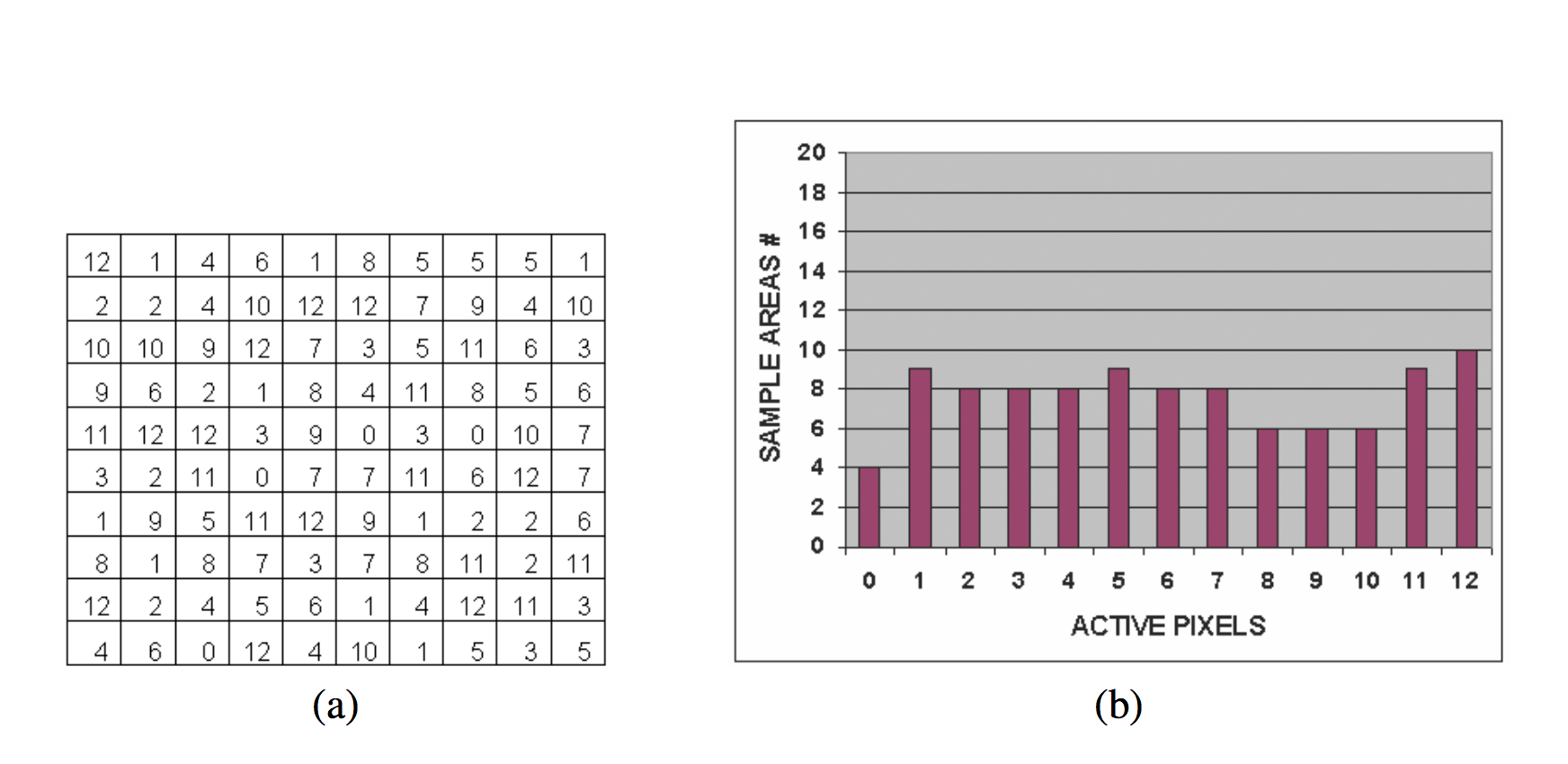}
\caption{\em Non uniform sampling}
\label{with-sampling}
\vspace{-5mm} \label{fig:graph}
\end{figure*}

%\begin{figure*}[t]
%\vskip 0mm
%\begin{center}
%\begin{tabular}{c@{\hspace{1cm}}c}
%  \resizebox{4.5cm}{!}{\includegraphics{repres-densities-c1.eps}} &
%  \resizebox{6.5cm}{!}{\includegraphics{repres-densities-c1-hist.eps}} \\
%  (a) & (b) \\
%\end{tabular}
%\caption{Non uniform sampling}
%\label{with-sampling}
%\vspace{-0mm}
%\end{center}
%\end{figure*}

The problem of representing relative densities is the one of
creating an optimal mapping between the set of the actual data
densities and the set of available represented densities. Each
data density must be associated to one of the 64 (under the
hypothesis of $8 \times 8$ sample areas) available represented
densities. Any given visualization is one particular mapping.
Consider the case in which a visualization is obtained by
displaying a large data set. It likely corresponds to a mapping in
which higher densities are all mapped onto few single represented
densities, the ones in which quite all pixels are active (pane
saturation). This is why in that areas relative densities can not
be perceived: a large number of high data densities is mapped onto
very close values. Our idea is to investigate how these mappings
could be changed in order to present to the user more information
about relative densities accepting, to a certain extent, some
distortion.

In the following we use a simple numeric example to clarify our
approach. Assume we are plotting 2264 (this strange number comes
from a random data generation) points on a screen composed by
400x400 pixels arranged in 100 sample areas of size 4x4 pixels. In
the example we concentrate on the number of data elements or
active pixels neglecting the SA area value (what we called A),
that is just a constant. In Figure~\ref{no-sampling}(a) the data
densities (in terms of number of points) corresponding to each
sample area are displayed.

Figure~\ref{no-sampling}(b) shows the actual values of data
densities (X axis) together with the associated number of sample
areas sharing each value (Y axis). As an example, we can see that
the maximum data density 49 is shared by just one sample area
($SA_{2,6}$) and the minimum data density  0 is shared by four
sample areas ($SA_{5,6}$,\ $SA_{5,8}$,\ $SA_{6,4}$,\ $SA_{10,3}$).
Figure~\ref{no-sampling} (c), obtained applying the statistical
results discussed in Section 2 (see Figure~\ref{fig:graph}), shows the actual represented
density (in terms of active pixels) ranging, for each $SA_{i,j}$, between 0 and 12. Looking at
figure~\ref{no-sampling} (d) it is easy to discover that more then
50\% of the visualization pane (54 sample areas out of 100)
ranging between 22 and 49 data density collapsed on just three
different represented data densities (10, 11, 12).

In order to improve such a situation we want to produce a new
mapping among the given data densities and the 12 available
represented densities. This can be done pursuing the goal of
preserving the maximum number of differences, loosing, on the
other end, their extent. In other words, we want to present the
user with as many difference in density as possible, partially
hiding the real amount of such differences.

In order to obtain such a result, starting from figure Figure~\ref{no-sampling} (b)
and considering only the 96 sample areas with data density $>0$ we
split the x axis in 12  (i.e., the available represented densities)
adjacent {\em non uniform} intervals, each of them containing
$96/12=8$ sample areas. Obviously, because of we are working on
discrete values we cannot guarantee that each interval contains
exactly 8 sample areas and we have to choose an approximation minimizing
the variance. After that, the data elements belonging to the sample areas
associated with the same interval $i$ are sampled in a way that
produces a represented density equal to $i$. As an example, the
first interval encompasses data densities 1 (shared by 6 sample
areas) and 2 (shared by 3 sample areas) and the associated data elements
are sampled as much as needed in order to produce a represented
density equal to 1. The second interval encompasses data density
3, 4, and 5 (7 sample areas) and, after the sampling, the
resulting data density is 2, and so on.

The represented densities resulting from this approach are
depicted in Figure~\ref{with-sampling} (a);
Figure~\ref{with-sampling} (b) shows the new, more uniform
distribution of such represented densities. We want to point out
that in this new representation the above collapsed 54 data
densities of Figure~\ref{no-sampling} (b) now range between 6 and
12 represented densities allowing the user to discover more
density differences. On the other hand, as an example, the real
difference between data densities 29 and 22 (1.32) is poorly
mapped on represented densities 7 and 6 (1.16).

Roughly speaking, we can think at the whole process as follows. We
have at disposal $p$ different represented densities that are
matched against $k$ real data densities where, usually, $k>>p$;
that implies that each represented density is in charge to
represent several, different data densities, hiding differences to
the user. The game is to change, by non uniform sampling, the
original data densities, altering their assignment to the $p$
available represented densities in order to preserve the
number of density differences.

\begin{figure*}[t]
\centering
\includegraphics[width=0.95\textwidth]{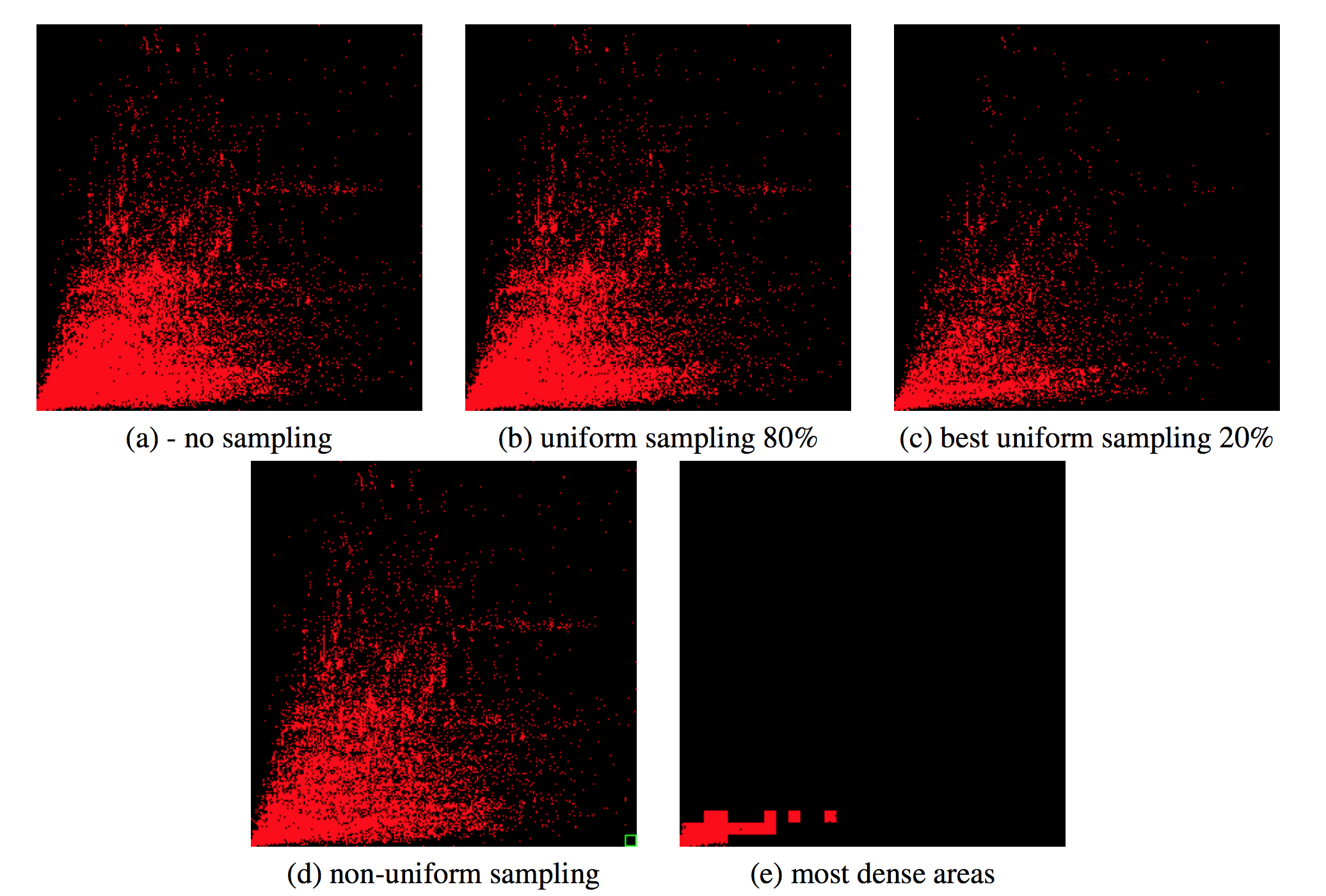}
\caption{Comparison of various sampling methods visualizing the
mail parcels dataset}
\label{fig:comp}
\vspace{-5mm} \label{fig:graph}
\end{figure*}

%\begin{figure*}[t]
%\vskip -10mm
%\begin{center}
%\begin{tabular}{c@{\hspace{0.5cm}}c@{\hspace{0.5cm}}c}
%  \resizebox{4.5cm}{!}{\includegraphics{original.eps}} &
%  \resizebox{4.5cm}{!}{\includegraphics{us-80.eps}} &
%  \resizebox{4.5cm}{!}{\includegraphics{us-20.eps}} \\
%  (a) - no sampling & (b) uniform sampling 80\% & (c) best uniform sampling
%  20\%\\
%\end{tabular}
%\begin{tabular}{c@{\hspace{0.5cm}}c}
%  \resizebox{4.5cm}{!}{\includegraphics{nus.eps}} &
%  \resizebox{4.5cm}{!}{\includegraphics{most-dense.eps}} \\
%  (d) non-uniform sampling & (e) most dense areas \\
%\end{tabular}
%\caption{Comparison of various sampling methods visualizing the
%mail parcels dataset} \label{fig:comp} \vspace{-0mm}
%\end{center}
%\end{figure*}

\section{Discussion}
\label{sec:discussion} In this section we show the effectiveness
of our technique commenting the images obtained applying different
sampling strategies. We compare the images acquired visualizing a
real dataset: the one containing 160,000 mail parcels already
mentioned in Section \ref{sec:modeling}.

The images come from a tool specifically developed for our
purposes. It is a Java based application that permits to inspect
several characteristics of the displayed image such as: the
data/represented density of each sample area, some quality metrics,
and the number of overlapping pixels. It is also  possible to
apply uniform and non-uniform sampling, and to filter sample areas
with data/represented density out of a specific range.

Figure \ref{fig:comp} shows: (a) the original visualization (no
sampling), (b) the one obtained uniformly sampling the data leaving 80\% of the
original dataset, (c) the one obtained uniformly sampling the data leaving 20\% of
the original dataset (this value is the best uniform sampling ratio computed by the proposal shown in \cite{BerSan04}), (d)the one obtained using non uniform
sampling. It is quite evident that a too weak uniform sampling
(Fig.\ref{fig:comp}(b)) does not make apparent density
differences in high density areas. Conversely, an optimized (but still too strong)
uniform sampling (Fig.\ref{fig:comp}(c)) makes them apparent
but to the detriment of low density areas. In fact, the upper
right area originally contained a cluster that is not visible
anymore. Figure \ref{fig:comp}(d) shows the result obtained when
applying non-uniform sampling. The features in the low density
areas are still visible (as in the case of weak uniform sampling)
but, at the same time, in the high density area it makes more
evident density differences that in the original image were not
perceptible (as in the case of strong uniform sampling). Figure
\ref{fig:comp}(e) makes it clearer. It is obtained filtering out the
sample areas with data density lower than 810 (i.e., SA with less
than 810 points) therefore showing the most dense areas. If
compared with the other images it is easy to notice that while on Figure \ref{fig:comp}(d), that pattern is perfectly clear, on  images (Fig. \ref{fig:comp}(a) and (b)) it is hidden in the saturated areas and on (Fig. \ref{fig:comp}(c) it is faintly visible. Roughly speaking, we
can say that our technique produce at the same time the advantages
of both strong and weak sampling.

Another interesting aspect worth to mention, is how this technique
can be operated. When applying uniform sampling, the choice of the
amount of sampling to choose is critical. If the sampling factor
is selected by hand, the user has to try many combinations until
s/he finds the value that best conveys the information. To overcome
this, we applied in \cite{BerSan04} an algorithm to
\emph{automatically} devise the amount of sampling to apply.
Exploiting the metrics presented in Section \ref{sec:metrics} we
were able to find the best sampling factor to apply, but the
problem of uniform sampling still held. Conversely, with
non-uniform sampling there is no need to search into a space of
solutions and the algorithm runs autonomously with the idea of
assigning the available represented densities as
\textit{smartly} as it can.

\begin{figure*}[t]
\vskip -20mm \centering
\includegraphics*[width=0.88\textwidth]{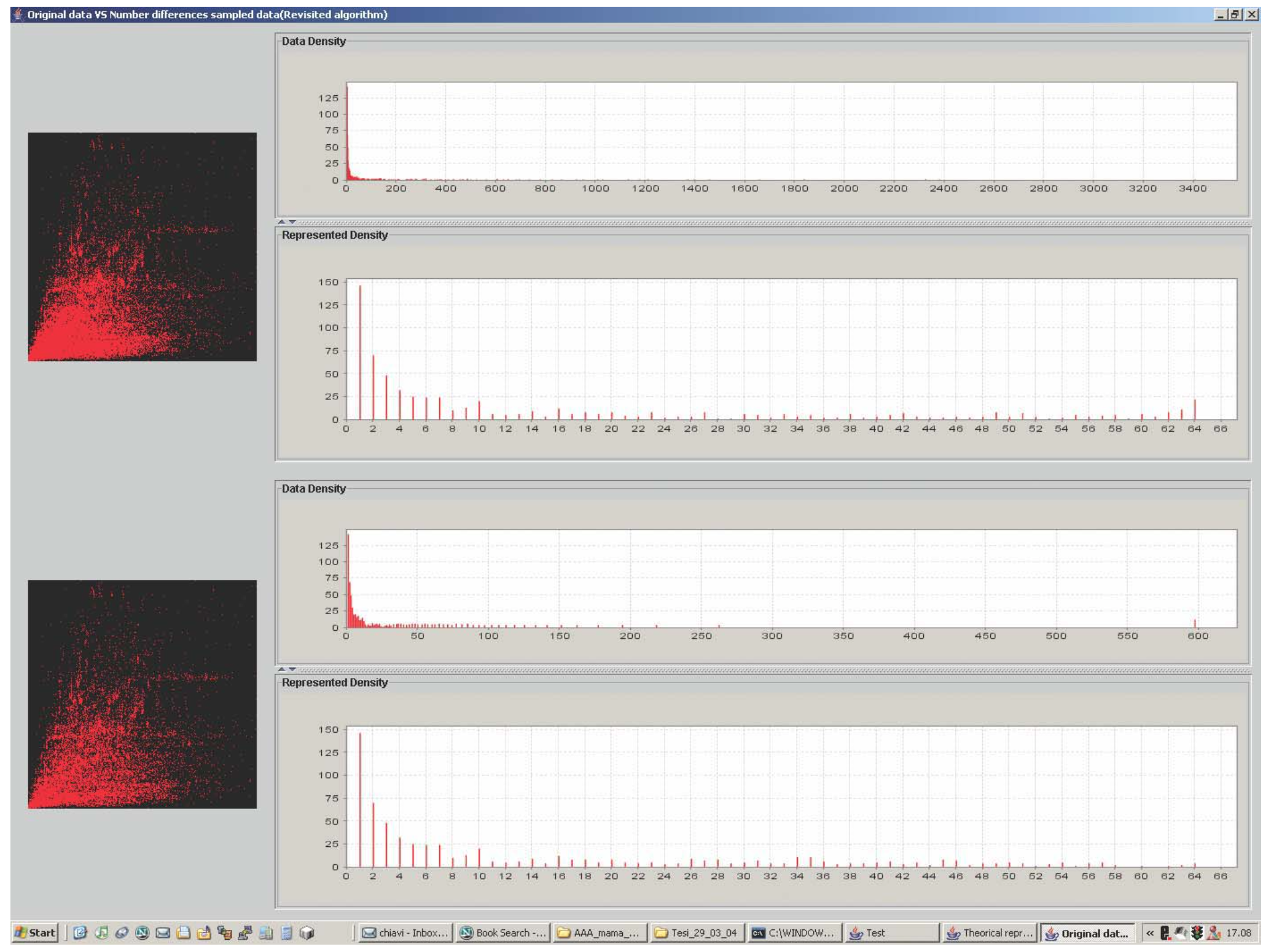}
\caption{{\em Non uniform sampling}} \vspace{-0mm}
\label{fig:nuni}
\end{figure*}

The logic behind the algorithm can be better appreciated
looking at Fig. \ref{fig:nuni} that compares the original and
non uniformly sampled visualizations together with their  densities histograms.
The densities are more evenly distributed, allowing the
dense areas to exhibit the underlying trends. Moreover, the peaks associated with the higher data densities (i.e., 62, 63, 64) are not present anymore.

\section{Conclusions and future work}
\label{sec:conclusions} In this paper we presented a low grain,
non uniform sampling sampling technique that automatically reduces
visual clutter in a 2D scatter plot and preserves relative
densities. To the best of our knowledge this approach is a quite
novel way of sampling visual data.
 The technique exploits some statistical results and a formal model describing
and measuring over plotting, screen occupation, and both data
density and represented data density. Such a model allows for computing \emph{where}, \emph{how}, and \emph{how much} to sample preserving  some image characteristics (i.e., relative density).

Several open issues rise from this work:
\begin{itemize}
\item users must be involved. Our strategy provides precise figures but
we need to map them against user perceptions. As an example, still
referring to our approach, if a sample area contains twice as many
active pixels as another one, does the user perceive the feeling
of observing a double density for any total occupation of the
areas? On the other hand, how much two sample areas may differ in
pixel number still giving the user the sensation of having the
same data density? We are currently designing some perceptive
experiments, in order to deep this aspect. The next step will be to incorporate within our algorithms these issues.
\item sampling areas. Several choices deserve more attention: it is
our intention to analyze the influence of increasing/decreasing of
sampling area dimension, in term of image quality and
computational aspects.
\end{itemize}

We are actually extending the prototype functionalities to apply
and verify our ideas. We want to implement a dataset generator  to
conduct controlled tests. The dataset generator will permit to
generate artificial distributions, giving the possibility
to control specific parameters, that will be used to create
specific cases considered critical or interesting.

\section{Acknowledgements}
We would like to thank Pasquale Di Tucci for his invaluable help in implementing the software prototype.
\bibliographystyle{plain}

\bibliography{IV_04}
%\end{multicols}
\end{document}